\documentclass[prd,aps,showpacs,tightenlines,nofootinbib,preprint,preprintnumbers]{revtex4}
\usepackage{graphicx}
\usepackage{amsfonts}
\usepackage{amssymb}
\usepackage{latexsym}

 \newcommand{\CL}{{\cal L}}

\newcommand{\bear}{\begin{array}}  \newcommand{\eear}{\end{array}}
\newcommand{\bea}{\begin{eqnarray}}  \newcommand{\eea}{\end{eqnarray}}
\newcommand{\beq}{\begin{equation}}  \newcommand{\eeq}{\end{equation}}
\newcommand{\bef}{\begin{figure}}  \newcommand{\eef}{\end{figure}}
\newcommand{\bec}{\begin{center}}  \newcommand{\eec}{\end{center}}
\newcommand{\non}{\nonumber}  
\newcommand{\lmk}{\left(}  \newcommand{\rmk}{\right)}
\newcommand{\lkk}{\left[}  \newcommand{\rkk}{\right]}
\newcommand{\lhk}{\left \{ }  \newcommand{\rhk}{\right \} }

\newcommand{\del}{\partial}  
\newcommand{\vect}[1]{\mbox{\boldmath${#1}$}}
\newcommand{\bib}{\bibitem}

\newcommand{\mg}{M_G}

\newcommand{\wv}{\widetilde\varphi}
\newcommand{\wc}{\widetilde\chi}
\newcommand{\dv}{\delta\varphi}
\newcommand{\dc}{\delta\chi}

\def\IBB#1#2#3{{\bf #1}, #2 (20#3)}

\def\APJSS#1#2#3{Astrophys. J., Suppl. Ser. {\bf #1}, #2 (20#3)}

\def\CQGG#1#2#3{Class. Quantum Grav. {\bf #1}, #2 (20#3)}

\def\JHEPP#1#2#3{J. High Energy Phys. {\bf #1}, #2 (20#3)}

\def\PLB#1#2#3{Phys. Lett. B {\bf #1}, #2 (19#3)}
\def\PLBB#1#2#3{Phys. Lett. B {\bf #1}, #2 (20#3)}
\def\PLBold#1#2#3{Phys. Lett. {\bf#1B}, #2 (19#3)}

\def\PRD#1#2#3{Phys. Rev. D {\bf #1}, #2 (19#3)}
\def\PRDD#1#2#3{Phys. Rev. D {\bf #1}, #2 (20#3)}

\def\PRL#1#2#3{Phys. Rev. Lett. {\bf#1}, #2 (19#3)}

\def\PRT#1#2#3{Phys. Rep. {\bf#1}, #2 (19#3)}

\def\PTPS#1#2#3{Prog. Theor. Phys. Suppl. {\bf #1}, #2 (19#3)}

\begin{document}

\title{Density fluctuations in one-field inflation}

\author{Masahide Yamaguchi} 
\affiliation{Department of Physics and Mathematics, Aoyama Gakuin
University, Kanagawa 229-8558, Japan}
\author{Jun'ichi Yokoyama} 
\affiliation{Research Center for the Early Universe (RESCEU), 
Graduate School of Science,
The University of
Tokyo, Tokyo, 113-0033, Japan}

\date{\today}
\preprint{RESCEU-39/05}


\begin{abstract}
Any one-field inflation is actually realized in a multifield
configuration because the inflaton must have couplings with other fields
to reheat the universe and is coupled to all other fields at least
gravitationally. In all single inflaton models, it is explicitly or
implicitly assumed that the heavier fields are stuck to their potential
minima during inflation, which are time-dependent in general. We present
a formalism to calculate curvature perturbations in such a
time-dependent background and show that the proper expression can be
obtained using a single-field analysis with a reduced potential in which
all these heavy fields are situated at their respective, time-dependent
minima. Our results provide a firm ground on the conventional
calculation.
\end{abstract}


\maketitle



\label{sec:intro}

We live in a big, spatially flat, globally homogeneous and isotropic
universe full of entropy today--- all by virtue of inflation in the
early universe \cite{inflation}.  The slow-roll inflation also generates
primordial density fluctuations as a seed for the large scale structure
of the universe \cite{pert}.  Recent observations of the cosmic
microwave background fluctuations by the WMAP \cite{WMAP,Peiris:2003ff}
and the BOOMERANG-03 \cite{Boomerang} found the concordance with the
predictions of inflation and confirmed that primordial density
fluctuations are adiabatic, Gaussian, and nearly scale-invariant.
Furthermore, the observation of the correlation between the temperature
anisotropy and the E-mode polarization has confirmed inflation-produced
adiabatic fluctuations and falsified causal seed models and primordially
isocurvature fluctuations as a generation mechanism of curvature
perturbation \cite{Peiris:2003ff}.

The simplest models of inflation require only one real scalar field,
which we call the {\it inflaton}, and inflation is driven as it slowly
rolls down along its potential toward a minimum.  The inflaton, however,
is by no means the sole scalar field of the theory because the inflaton
must have couplings with other fields to reheat the universe and is
coupled to all other fields at least gravitationally. Furthermore, a lot
of inflation models are recently considered in the stringy landscape
\cite{landscape}, in which there are a large number of scalar fields
with a huge number of extrema. We simply ignore the existence of the
other fields by postulating either that they are anchored at their
respective minima during inflation due to some potential, or that their
energy scale is so low that neither their homogeneous parts nor their
fluctuations play important roles. Thus we should always keep in mind
that inflation occurs in the multifield configuration in a sensible
theory of high energy physics.

A number of analyses have been done on inflation models with multiple
scalar fields, which involve many fluctuating fields, focused on the
generation of isocurvature fluctuations in addition to the adiabatic
fluctuations \cite{iso}.  To our knowledge, however, the other important
class of inflation models with multiple scalar fields mentioned above
has not been fully investigated yet, in which only one field $\varphi$
is light and the others $\chi_i\,(i = 1, \dots, N)$ are heavy. In such a
situation, the light field $\varphi$ plays a role of the inflaton but
the other heavy fields $\chi_i$ may affect the trajectory of the light
field. In fact, virtually all the single inflaton models based on modern
unified theories fall into this category for the reason stated above,
either manifestly or implicitly. In particular, inflation models are
often considered in the stringy landscape, in which there are a large
number of scalar fields with a huge number of extrema. In the folded
inflation model, for example, the inflaton turns several corners in the
potential \cite{folded}.

Among the manifestly multifield models, the best known example is the
hybrid inflation model proposed by Linde \cite{hybrid}, which is
recently paid particular attention to as a model to realize the running
feature of the spectral index of density fluctuations inferred by the
WMAP results \cite{running}. In hybrid inflation, the heavy fields are
stuck on the origin $\chi_i=0$ due to their heavy masses during
inflation so that the light field $\varphi$ rolls down along the
effective potential $V(\varphi) = V(\varphi, \chi_i=0)$. Thus, we look
on hybrid inflation as an effectively single field inflation and
calculate primordial density perturbations by use of the effective
potential $V(\varphi) = V(\varphi, \chi_i=0)$.  The original hybrid
inflation, however, is exceptional among the manifestly multifield
models because the light and the heavy fields are orthogonal to each
other in the sense that the minima $\chi_i^m = 0$ are independent of the
light field.

We must point out, however, that in general the minima of heavy fields
are dependent on the inflaton. This is because all the scalar fields are
coupled with each other at least with the gravitational strength.  Then
the expectation values of other heavy fields may well shift
significantly during inflation, being affected by the motion of the
inflaton. For example, the number of $e$-folds elapsed while the
inflaton changes by $\Delta\phi$ is given by
\beq 
  N=8\pi G \left| \int^{\phi}_{\phi-\Delta\phi}
             \frac{V(\phi)}{V'(\phi)} d\phi
           \right|
 \approx \frac{(\Delta\phi)^2}{\mg^2}, 
\eeq 
for a typical polynomial type potential of the inflaton $V(\phi)$ which
is used to realize chaotic inflation \cite{chaoinf}.  Thus during the
last 55 $e$-folds corresponding to the observable regime, the inflaton
changes more than the reduced Planck scale $\mg=2.4\times
10^{18}$GeV. This is also the case with many small-field models such as
new \cite{newinf} and topological inflation models \cite{topoinf}
because the vacuum expectation value after inflation typically takes a
value of order of $\mg$. Therefore, this change would inevitably affect
the behavior of other heavy fields even if they are coupled with the
inflaton only through gravitational strength, whose back reaction to the
inflaton may not be negligible in turn.

Furthermore in some types of small-field inflation models, in which
coupling constants of interactions of heavy fields are much less than
unity, the potential minima of these heavy fields can be modulated
considerably thorough gravitationally suppressed interactions, even when
change of the inflaton is much {\it smaller} than $\mg$.  A good example
is the smooth hybrid inflation proposed by Lazarides and
Panagiotakopoulos \cite{smooth}. In this model, the potential for a
light field $\varphi$, which plays the role of the inflaton, and a heavy
field $\chi$ is given by
\beq
  V(\varphi,\chi) = \lmk \mu^2 - \frac{\chi^4}{16 M^2} \rmk^2
                     + \frac{\chi^6 \varphi^2}{16 M^4},
\eeq 
where $\mu$ and $M$ are some mass scales.
For $\varphi^2 \gg \mu M$, the potential minimum of $\chi$ is given by
\beq
  (\chi^m)^2 \simeq \frac43 \frac{\mu^2 M^2}{\varphi^2}.
\eeq
Note that $\chi^m$ is dependent on $\varphi$ here, that is, $\chi^m =
\chi^m(\varphi)$. Since the effective mass squared 
 at the minimum $\chi^m$, $m_{\chi}^2 \simeq
4\mu^4 / (3\varphi^2)$, is much larger than the
Hubble parameter squared, $H^2 \simeq \mu^4 / (3 M_G^2)$, as long as
$\varphi \ll M_G$, $\chi$ quickly relaxes at the minimum during
inflation. Then, the dynamics of the homogeneous mode of the $\varphi$
field is described by the effective single-field potential $V_{\rm
s}(\varphi) = V(\varphi, \chi = \chi^m(\varphi))$ and leads to
inflation. Recently, this type of inflation is found to be attractive
\cite{YY,KTYY} because, combined with new inflation \cite{newinf}, it can
simultaneously explain the two new features discovered by the recent
precision measurements of cosmic microwave background anisotropy, that
is, the running of spectral index of density fluctuations on large scale
as preferred by the first-year WMAP data and the BOOMERANG-03, and a
large enough amplitude of fluctuation on small scale relevant to first
star formation to realize early re-ionization as discovered by WMAP.

In \cite{YY}, however, primordial density fluctuations have been
calculated by looking on such models as  effectively single-field
inflation of the inflaton $\varphi$ with the potential $V_{\rm
s}(\varphi) = V(\varphi, \chi = \chi^m(\varphi))$. Although it is true
that the dynamics of the homogeneous mode is determined by the effective
single-field potential, it is a nontrivial task to justify the above
procedure to calculate primordial density fluctuations, which enables us
to estimate the effects of the change of the minima of heavy fields on
density fluctuations for the first time.

As mentioned before, the existence of the other fields is often ignored
simply postulating either that they are so heavy that they are anchored
at their respective minima during inflation, or that their energy scale
is so low that neither their homogeneous parts nor their fluctuations
play important roles. However, the considerations about the mass scale
or the energy scale of heavy fields are not sufficient to estimate their
effects on density fluctuations. That is, even if changes of the minima
of heavy fields are very small, there is no guarantee that such changes
do not have significant effects on density fluctuations. Indeed they
depend not only on the potential energy density, which is hardly
affected by the small change of heavy fields, but also the kinetic term,
for their amplitude is inversely proportional to the time derivative of
the inflaton.  Since its kinetic energy is much smaller than the
potential energy during inflation, a small shift in the heavy field may
well induce a large relative change in the kinetic term, which may
affect the amplitude of density fluctuation considerably.

In this paper, we investigate this problem in detail. That is, we
consider inflation models with multiple scalar fields, in which only one
field $\varphi$ is light and the others $\chi_i ~(i = 1, \dots, N)$ are
heavy. During inflation, the heavy fields $\chi_i$ are stuck to their
minima $\chi_i^m(\varphi)$, which can depend on the light field
$\varphi$. We study creation of perturbations in this setting. Since
there is only one light mode that acquires long-wave quantum
fluctuations during inflation, only the adiabatic fluctuation is
important. We derive the relevant equations and compare with the
conventional approach of the single-field approximation.  As a result we
show that this approximation is completely justified.


\label{sec:DF}

First, we consider inflation models in multifield configurations, $\wv$
and $\wc_i ~(i = 1, \dots N)$. The Lagrangian density is given by
\beq
  \CL =  - \frac12 (\del_{\mu} \wv)^2 - \frac12 (\del_{\mu} \wc_i)^2
         - V(\wv,\wc_i).
\eeq
Here we assume that the field $\wv$ is light and the other fields
$\wc_i$ are heavy during inflation, that is,
\beq
   |V_{,\wv\wv}| \ll H^2
   \qquad {\rm and} \qquad
   V_{,\wc_i \wc_i} \gg H^2,
\label{eq:masses}
\eeq
where $H \equiv \dot{a}/a$ is the Hubble parameter during inflation, a
dot represents derivative with respect to the cosmic time $t$, and a
comma represents partial derivative with respect to the corresponding
field variables. Hereafter we set $N = 1$ and denote $\wc_1$ by $\wc$
for simplicity. The extension to the case with $N \ge 2$ is trivial.

We decompose the two fields to homogeneous modes and perturbations,
\bea
  && \wv(t,\vect{x}) = \varphi(t) + \dv(t,\vect{x}),
     \non \\
  && \wc(t,\vect{x}) = \chi(t) + \dc(t,\vect{x}).
\eea
The perturbed metric around the flat Friedmann universe in the
longitudinal gauge is taken as
\beq
  ds^2 = -\lkk 1+2\Phi(t,\vect{x})\rkk dt^2 
                 + a(t)^2\,\lkk 1-2\Phi(t,\vect{x})\rkk
  \delta_{ij}dx^i dx^j
\eeq 
with $\Phi(t,\vect{x})$ being  the gravitational potential. 

First of all, we consider the dynamics of the homogeneous modes
$\varphi(t)$ and $\chi(t)$. The equations of motion for the
homogeneous modes are given by
\bea
  && \ddot{\varphi} + 3H \dot{\varphi} 
       + V_{,\varphi} = 0, \non \\
  && \ddot{\chi} + 3 H \dot{\chi} 
       + V_{,\chi} = 0, \\
  && H^2 =\frac{1}{3M_G^2}
\lmk\frac{1}{2}\dot\varphi^2+\frac{1}{2}\dot\chi^2+V\rmk. \non
\eea
Since the mass of $\chi$ is much larger than the Hubble parameter, the
field $\chi$ quickly relaxes to the minimum $\chi^m = \chi^m(\varphi)$,
characterized by
\beq
  V_{,\chi}(\varphi,\chi^m(\varphi)) = 0.
\eeq
Since this relation holds for arbitrary $\varphi$, we find
\beq
  \frac{dV_{,\chi}}{d\varphi}(\varphi,\chi^m(\varphi))
    = V_{,\chi\varphi} 
     + V_{,\chi\chi} \frac{d\chi^m}{d\varphi}
    = 0.
  \label{eq:relation} 
\eeq
Defining the effective single-field potential $V_{\rm s}(\varphi)$ as
$V_{\rm s}(\varphi) = V(\varphi,\chi = \chi^m(\varphi))$, the equation
of motion for $\varphi$ reads
\beq
  \ddot{\varphi} + 3 H \dot{\varphi} 
       + \frac{dV_{\rm s}}{d\varphi} = 0,
  \label{eq:homo}
\eeq
where we have used 
\beq
\frac{dV_{\rm s}}{d\varphi}(\varphi,\chi^m(\varphi)) =
\left.\frac{\partial V}{\partial \varphi}\right|_{\chi=\chi^m(\varphi)}
+\frac{d\chi^m(\varphi)}{d\varphi}
\left.\frac{\partial V}{\partial \chi}\right|_{\chi=\chi^m(\varphi)}
=
V_{,\varphi}(\varphi,\chi^m(\varphi)).
\eeq
Thus, the dynamics of the homogeneous mode $\varphi$ is completely
determined by the reduced potential $V_{\rm s}(\varphi)$. That is, we
can look on this type of inflation as an effectively single-field
inflation with the potential $V_{\rm s}(\varphi)$, as far as the
dynamics of the homogeneous mode is concerned. 

Note that for successful inflation, the absolute magnitude of the
effective mass squared of $\varphi$ must be much smaller than the Hubble
parameter squared, that is,
\beq
  \left| \frac{d^2 V_{\rm s}}{d\varphi^2} \right|
   = \left| V_{,\varphi\varphi} 
      + 2 \frac{d\chi^m}{d\varphi} V_{,\varphi\chi}
      + \lmk \frac{d\chi^m}{d\varphi} \rmk^2 V_{,\chi\chi} \right|
   = \left| \frac{V_{,\varphi\varphi}V_{,\chi\chi}-V_{,\varphi\chi}^2}
          {V_{,\chi\chi}} \right|
   \ll H^2,
  \label{eq:ems}
\eeq
where we have used Eq. (\ref{eq:relation}). One sufficient condition for
this inequality is that the non-diagonal components of Hesse matrix of
the potential $V$ are not so large, that is,
\beq
  V_{\,,\varphi\chi}^2 \lesssim 
   |V_{\,,\varphi\varphi}\,V_{\,,\chi\chi}|.
\label{eq:nondiagonal}
\eeq

Next, we consider  perturbations of the field variables, $\delta\varphi$
and $\delta\chi$, whose  equations of motion are given by
\cite{fluc}
\bea
  && \ddot{\dv} + 3 H \dot{\dv} - \frac{\nabla^2}{a^2} \dv + 
         V_{,\varphi\varphi} \dv
      = -2 V_{,\varphi} \Phi + 4 \dot{\varphi}\dot{\Phi} 
         - V_{,\varphi\chi} \dc, \non \\
  && \ddot{\dc} + 3 H \dot{\dc} - \frac{\nabla^2}{a^2} \dc + 
         V_{,\chi\chi} \dc
      = -2 V_{,\chi} \Phi + 4 \dot{\chi} \dot{\Phi} 
         - V_{,\chi\varphi} \dv.
\eea
On the other hand, the evolution of the gravitational potential is
governed by
\bea
  && \dot{\Phi} + H \Phi = \frac{1}{2M_G^2} 
       (\dot{\varphi} \dv + \dot{\chi} \dc), \non \\
  && \ddot{\Phi} + H \dot{\Phi} + \frac{\nabla^2}{a^2} \Phi
       = \frac{1}{M_G^2} 
       (\dot{\varphi} \dot{\dv} + \dot{\chi} \dot{\dc}).
\eea

We are interested only in the adiabatic density fluctuations, which is
characterized by the condition \cite{iso},
\beq
  \frac{\dv}{\dot{\varphi}} = \frac{\dc}{\dot{\chi}},
  \label{eq:ad}
\eeq
or equivalently
\beq
  \dc = \frac{d\chi^m(\varphi)}{d\varphi} \dv,
  \label{eq:ad2}
\eeq
where we have used
$\dot{\chi}/\dot{\varphi}=d\chi^m(\varphi)/d\varphi$. Inserting the
relation (\ref{eq:relation}), the adiabatic condition (\ref{eq:ad2}) can
be rewritten as
\beq
  \dc = \frac{d\chi^m}{d\varphi} \dv
      = - \frac{V_{,\chi\varphi}}{V_{,\chi\chi}} \dv.
  \label{eq:ad3}
\eeq
Under this condition, the equation of the perturbation $\dv$ becomes
\beq
  \ddot{\dv} + 3 H \dot{\dv} - \frac{\nabla^2}{a^2} \dv 
    + \lmk 
          \frac{V_{,\varphi\varphi}V_{,\chi\chi}-V_{,\varphi\chi}^2}
               {V_{,\chi\chi}}                         
          \rmk \dv
      = -2 V_{,\varphi} \Phi + 4 \dot{\varphi}\dot{\Phi}.
  \label{eq:dv}
\eeq
Here note that $\dv$ is still light, that is, the absolute magnitude of
the effective mass squared of $\dv$ is also much smaller than $H^2$ due
to Eq. (\ref{eq:ems}).
 
The equations of the gravitational potential are then written as
\bea
  &&
  \dot{\Phi} + H \Phi = \frac{1}{2M_G^2}
   \lkk 1 + \lmk \frac{d\chi^m}{d\varphi} \rmk^2 \rkk 
     \dot{\varphi} \dv, \\
  && \ddot{\Phi} + H \dot{\Phi} + \frac{\nabla^2}{a^2} \Phi 
    = \frac{1}{M_G^2} 
       \lhk
         \lkk 1 + \lmk \frac{d\chi^m}{d\varphi} \rmk^2 \rkk    
            \dot{\varphi} \dot{\dv}
         + \frac{d^2\chi^m}{d\varphi^2} \frac{d\chi^m}{d\varphi}
            \dot{\varphi}^2 \dv
       \rhk.
  \label{eq:gp}
\eea
From these equations, the gravitational potential can be described by
$\dv$,
\bea
 \lmk \dot{H} - \frac{\nabla^2}{a^2} \rmk \Phi
    &=& \frac{1}{2M_G^2} 
       (\ddot{\varphi} \dv - \dot{\varphi} \dot{\dv} 
        + \ddot{\chi} \dc - \dot{\chi} \dot{\dc}), \non \\
    &=& \frac{1}{2M_G^2} 
       (\ddot{\varphi} \dv - \dot{\varphi} \dot{\dv}) 
       \lkk 1 + \lmk \frac{d\chi^m}{d\varphi} \rmk^2 \rkk.
 \label{eq:gpdirect} 
\eea 
Here $\dot{H}$ is given by
\beq
  \dot{H} = - \frac{1}{2M_G^2} (\dot{\varphi}^2 + \dot{\chi}^2)
          = - \frac{1}{2M_G^2} \dot{\varphi}^2
              \lkk 1 + \lmk \frac{d\chi^m}{d\varphi} \rmk^2 \rkk.
\eeq 
Therefore, in the long wave limit, the gravitational potential is given
by
\beq
  \Phi =  \frac{d~}{dt}\lmk \frac{\dv}{\dot{\varphi}}\rmk.
  \label{eq:gravitationalpot}
\eeq
Thus, adiabatic fluctuations produced in this inflation model are
completely characterized by equations (\ref{eq:homo}), (\ref{eq:dv}),
and (\ref{eq:gravitationalpot}).

Next, we investigate whether the above two equations can be
reproduced by considering a single-field inflation model
with the effective
potential $V_{\rm s}(\varphi) = V(\varphi,\chi^m(\varphi))$. For this
purpose, let us consider the following Lagrangian density,
\beq
  \CL =  - \frac12 (\del_{\mu} \wv)^2 - V_{\rm s}(\wv).
\eeq
The equation of motion for the homogeneous mode $\varphi$ is given by
the same equation (\ref{eq:homo}). The equation of motion of $\dv$ for
this system is given by
\beq
  \ddot{\dv} + 3 H \dot{\dv} - \frac{\nabla^2}{a^2} \dv + 
     \frac{d^2 V_{\rm s}}{d\varphi^2} \dv
      = - 2 \frac{d V_{\rm s}}{d\varphi} \Phi 
          + 4 \dot{\varphi}\dot{\Phi}. 
  \label{eq:dv2}
\eeq
The derivative of the effective potential $V_{\rm s}(\varphi) =
V(\varphi,\chi^m(\varphi))$ can be described by the partial derivative
of the original potential,
\bea
  && \frac{d V_{\rm s}}{d\varphi} 
    = V_{,\varphi}+V_{,\chi}\frac{d\chi^m}{d\varphi}=V_{,\varphi}, \non \\
  && \frac{d^2 V_{\rm s}}{d\varphi^2} 
   =  V_{,\varphi\varphi} 
      + 2 \frac{d\chi^m}{d\varphi} V_{,\varphi\chi}
      + \lmk \frac{d\chi^m}{d\varphi} \rmk^2 V_{,\chi\chi}
   = \frac{V_{,\varphi\varphi}V_{,\chi\chi}-V_{,\varphi\chi}^2}
          {V_{,\chi\chi}}.
\eea
Inserting this relation to Eq. (\ref{eq:dv2}) yields
\beq
  \ddot{\dv} + 3 H \dot{\dv} - \frac{\nabla^2}{a^2} \dv + 
    \lmk 
          \frac{V_{,\varphi\varphi}V_{,\chi\chi}-V_{,\varphi\chi}^2}
          {V_{,\chi\chi}} 
        \rmk \dv
      = -2 V_{,\varphi} \Phi + 4 \dot{\varphi} \dot{\Phi},
\eeq
which completely coincides with Eq. (\ref{eq:dv}). 

The equations of the gravitational potential are written as
\bea
  &&  \dot{\Phi} + H \Phi = \frac{1}{2M_G^2} \dot{\varphi} \dv, \\
  && \ddot{\Phi} + H \dot{\Phi} + \frac{\nabla^2}{a^2} \Phi
       = \frac{1}{M_G^2} t{\varphi} \dot{\dv}.
  \label{eq:gp2}
\eea
From these equations, the gravitational potential can be described by
$\dv$,
\beq
 \lmk \dot{H} - \frac{\nabla^2}{a^2} \rmk \Phi
    = \frac{1}{2M_G^2} 
       (\ddot{\varphi} \dv - \dot{\varphi} \dot{\dv}).
 \label{eq:gpdirect2} 
\eeq 
Note that this equation is different from the equation
(\ref{eq:gpdirect}) by the factor $[ 1 + (d\chi^m/d\varphi)^2 ]$.
However, here $\dot{H}$ is given by
\beq
  \dot{H} = - \frac{1}{2M_G^2} \dot{\varphi}^2.
\eeq 
Therefore, in the long wave limit, the gravitational potential is given
by
\beq
  \Phi = \frac{d~}{dt}\lmk \frac{\dv}{\dot{\varphi}}\rmk ,
  \label{eq:gravitationalpot2}
\eeq
which completely coincides with Eq. (\ref{eq:gravitationalpot}).

Thus, this type of inflation can be regarded as an effectively 
single-field inflation with the reduced potential 
$V_{\rm s}(\varphi)$ not
only for the dynamics of the homogeneous mode but also for generation of
adiabatic density fluctuations.

Finally we comment on the physical meaning of the adiabatic condition
(\ref{eq:ad}) in our system. The homogeneous mode of the heavy field is
assumed to stay at its minimum during inflation, that is,
\beq
  V_{,\chi}(\varphi,\chi) = 0, \quad
  {\rm namely,} \quad
  \chi = \chi^m(\varphi).
\eeq
Under this condition, we can show that
 the adiabatic condition (\ref{eq:ad}) is equivalent to the
following requirement,
\beq
  V_{,\chi}(\wv,\wc) = 0, \quad
  {\rm namely,} \quad
  \wc = \chi^m(\wv).
\eeq
That is, $\wc$ is related to $\wv$ through the same function $\chi^m$
as the case of the homogeneous mode and the perturbed fields also
stay at the minimum of $\chi$ for adiabatic fluctuations. This can be easily
shown. First, we expand $V_{,\chi}(\wv,\wc)$ as
\beq
  V_{,\chi}(\wv,\wc) = V_{,\chi}(\varphi+\dv,\chi+\dc)
                     = V_{,\chi\varphi}(\varphi,\chi)\dv
 + V_{,\chi\chi}(\varphi,\chi)\dc
                     = V_{,\chi\chi}(\varphi,\chi) \lmk 
                         - \frac{d\chi^m}{d\varphi} \dv + \dc \rmk,
\eeq
where we have used the relation (\ref{eq:relation}) in the last equation.
Then, it is obvious that $V_{,\chi}(\wv,\wc) = 0$ is equivalent to
\beq
  \dc = \frac{d\chi^m}{d\varphi} \dv 
      = \frac{\dot{\chi}}{\dot{\varphi}} \dv,
\eeq
which is just the adiabatic condition (\ref{eq:ad}).


To conclude, we have discussed primordial density fluctuations
produced by an inflation model with multiple scalar fields in which only
one field is light and the others are heavy. We have shown that
adiabatic density fluctuations of such an inflation can be completely
reproduced by looking on it as an effectively single-field model with
the reduced potential obtained by inserting the minima of the heavy
fields to the original potential.

\acknowledgments{ We are grateful to Robert H. Brandenberger for his
useful comments. M.Y.\ is supported in part by the project of the
Research Institute of Aoyama Gakuin University. J.Y.\ was partially
supported by the JSPS Grant-in-Aid for Scientific Research No.\
16340076.}

\end{document}